\documentclass[a4paper,11pt]{article}
\pdfoutput=1

\usepackage{jinstpub}
\usepackage{lineno}
\graphicspath{{figure/}}

\title{The TPC detector of PandaX-III Neutrinoless Double Beta Decay experiment}
\author[a,b]{Shaobo Wang (On behalf of PandaX-III collaboration)} 

\affiliation[a]{INPAC and School of Physics and Astronomy, Shanghai Jiao Tong University, \\Shanghai Laboratory for Particle Physics and Cosmology, Shanghai 200240, China}
\affiliation[b]{ SPEIT (SJTU-ParisTech Elite Institute of Technology), Shanghai Jiao Tong University, Shanghai 200240, China}

\emailAdd{shaobo.wang@sjtu.edu.cn}

\abstract{
The PandaX-III experiment uses a high pressure xenon gas TPC to search for the NLDBD of $^{136}$Xe. 
 The Microbulk Micromegas will be used for charge amplification and readout in order to record both the energy and track of an event with high energy resolution and spatial resolution.
 In the first phase of the experiment, we are building a detector which contains 140 kg 90\% $^{136}$Xe enriched gas operated at 10 bar. The detector will be installed in a dry shielding system at CJPL-II.
In this paper, we wil report  the design of the PandaX-III detector,  including the shielding of the experiment and the components of the TPC: the high pressure vessel, the readout plane and the electric field cage.
}

\arxivnumber{2001.01356}

\begin{document}
\maketitle
\flushbottom

\section{Introduction}
\label{sec:intro}

TPC (Time Projection Chamber) technology is widely used in particle physics experiments since the invention in 1970s~\cite{Nygren:1978rx}.
More recently, TPC’s application is expanded to rare event searches~\cite{Irastorza:2015dcb}, for example, in NLDBD~(Neutrinoless Double Beta Decay) searches, in the direct detection of WIMPs~\cite{Iguaz:2015myh}, axions and other WISPs~\cite{Alquezar:2013svd}. 
The requirements of these rare event search experiments are demanding, essentially a very sensitive detector with low radioactivity. 
The availability of a powerful description of the events recorded by the detector is important for the discrimination of the signal from the background.  
Nowadays many international collaborations are searching for NLDBD of  $^{136}$Xe with the  high pressure xenon gas TPC, e.g.  NEXT~\cite{Martin-Albo:2015rhw} and PandaX-III~\cite{Chen:2016qcd}.  
Gaseous TPC can record both the 3D spatial coordinates and the energy depositions $dE/dx$ for every track sampling point precisely, which are powerful sources of information about the event.
The generic requirements for a gaseous TPC intended for rare event searches are very good imaging capabilities and energy resolution, high gain and efficiency,  stability and radiopurity.
 Especially for the PandaX-III experiment, such requirements could be fulfilled by TPCs,  because they are equipped with Microbulk Micromegas~(MMs)~\cite{Andriamonje:2010zz}.
 The Microbulk Micromegas is made of Kapton and copper, so it guarantees extremely good radiopurity levels.
 The mechanical homogeneity of the gap and mesh geometry is superior, and thus the energy resolution is excellent~\cite{Iguaz:2012ur,Gonzalez-Diaz:2015oba}.  
 
 The PandaX-III experiment will construct a TPC which can hold 140~kg of xenon at 10~bar operating pressure. 
 The detector has an active volume with a 120~cm drift distance and a 160~cm diameter.
 The charge readout plane consists of a tessellation of 52 20$\times$20~cm$^2$ Microbulk Micromegas readout modules.
 The electric field cage of the TPC is designed with the flexible PCB connected with SMD resistors, and supported by an acrylic barrel. 
 The copper substrate inside the vessel and the Pb/HDPE layers outside the detector are designed in order to  get an excellent control over background.
 
The structure of this paper is as follows.
Firstly the overview of the design of PandaX-III experiment will be presented in Section~\ref{sec:Overview}.
Then I will detail the design of the TPC itself in Section~\ref{sec:TPC}, such as the vessel, the charge readout plane and the electric field cage.
Finally I give the conclusions in Section~\ref{sec:outlook}.

\section{Overview of the design of PandaX-III experiment}
\label{sec:Overview}
\begin{figure}[tb]
  \begin{center}
    \includegraphics[width=0.6\linewidth]{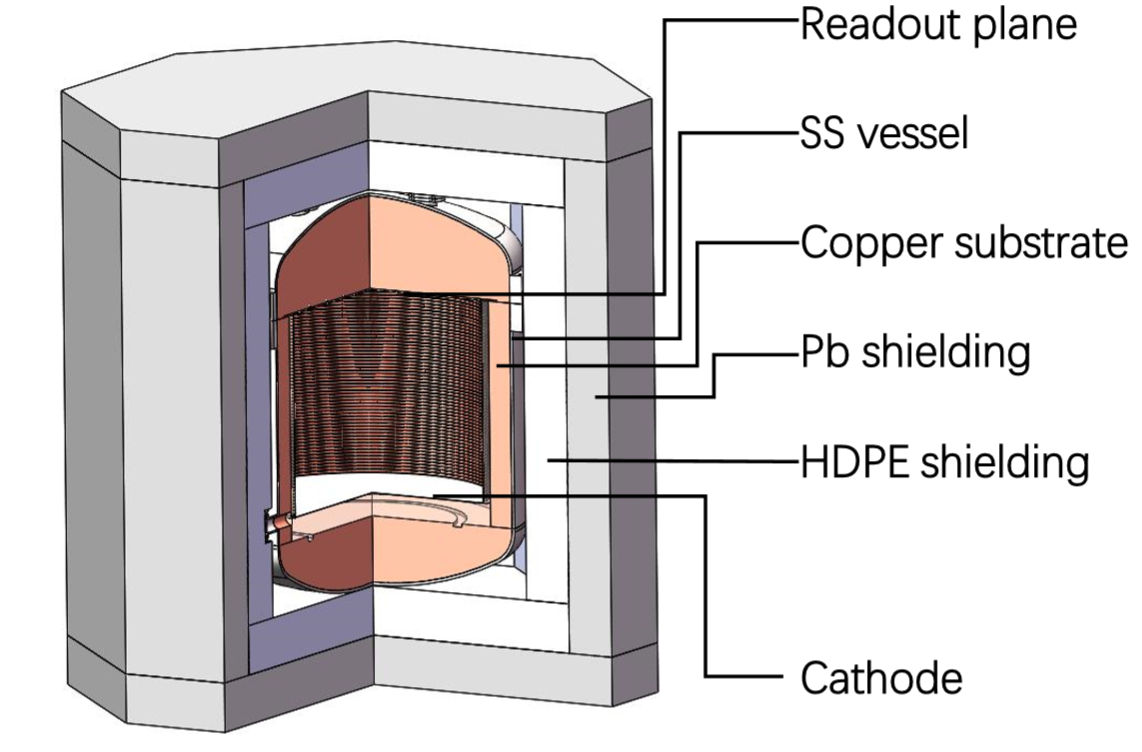}
\caption{Illustration of the PandaX-III experimental setup from the dry shielding system to the TPC itself, and from the cathode on the bottom to the charge readout plane with Micromegas on the upper side. All the main components  are annotated.}
   \label{fig:shielding}
 \end{center}
\end{figure}

The overview of the PandaX-III setup is shown in Fig.~\ref{fig:shielding}.
The central part of  the experimental setup is a high pressure xenon gas TPC.
It consists of a charge readout plane on the top, a cathode at the bottom and an electric field shaping cage in between.
The active volume of the TPC is of cylindrical shape with a height of 120.0~cm, a diameter of 160.0~cm. The charge readout plane is hanging from the top flat flange of a stainless steel (SS) pressure vessel, which could be operated up to 10~bar.
The cathode and the electric field shaping cage are installed inside the copper substrate with a width of 15~cm, which is further used to shield the radioactive background outside of the TPC, and also used to replace the $^{136}$Xe gas. 
The detector is surrounded by a dry shielding system, which consists of a layer of Pb with a width of 30.0~cm and a layer of HPDE with a width of 30.0~cm.
The setup will be installed at the CJPL-II ( China Jin-Ping underground Laboratory II),  
with the rock overburden of 2400~m, the rate of cosmic muons is reduced to 1/m$^2$/week.

While constructing the TPC, we paid special attention to the design and fabrication of individual parts,  as well as material selection to fulfill the requirement of vacuum and high gas pressure up to 10~bar.
On the other hand, we also enforced the radioactive contamination screening of the low background stainless steel, which is used in PandaX~\cite{Tan:2016zwf} dark matter experiment. 
And we have simulated the background budget for this design, which is comparable to the results in our CDR~\cite{Chen:2016qcd}. For the performance of PandaX-III experiment, with a projected energy resolution of 3\% FWHM, a signal efficiency of 35\%, 
and a background rate of 10$^{-4}$ c/keV/kg/yr after topological cuts, the PandaX-III  140~kg level experiment reaches a half-life sensitivity (90\% CL) of 8.5$\times$10$^{25}$ years after 3 years of live time.

\section{Time Projection Chamber}
\label{sec:TPC}
The PandaX-III detector is a high pressure xenon gas TPC which contains 140~kg $^{136}$Xe (90\% enriched) in the active volume inside the electric field cage. The working gas is 10~bar Xenon(99\%):TMA(1\%)~\cite{Chen:2016qcd}. 
A copper substrate with a width of 15.0~cm is placed between the SS vessel and the TPC. 
The detector is single-end readout with 52 20$\times$20~cm$^2$ Microbulk Micromegas.
The design and construction of each components are described below.

\subsection{High SS pressure vessel}
\label{sec: vessel}

\begin{figure}[tb]
  \begin{center}
    \includegraphics[width=0.8\linewidth]{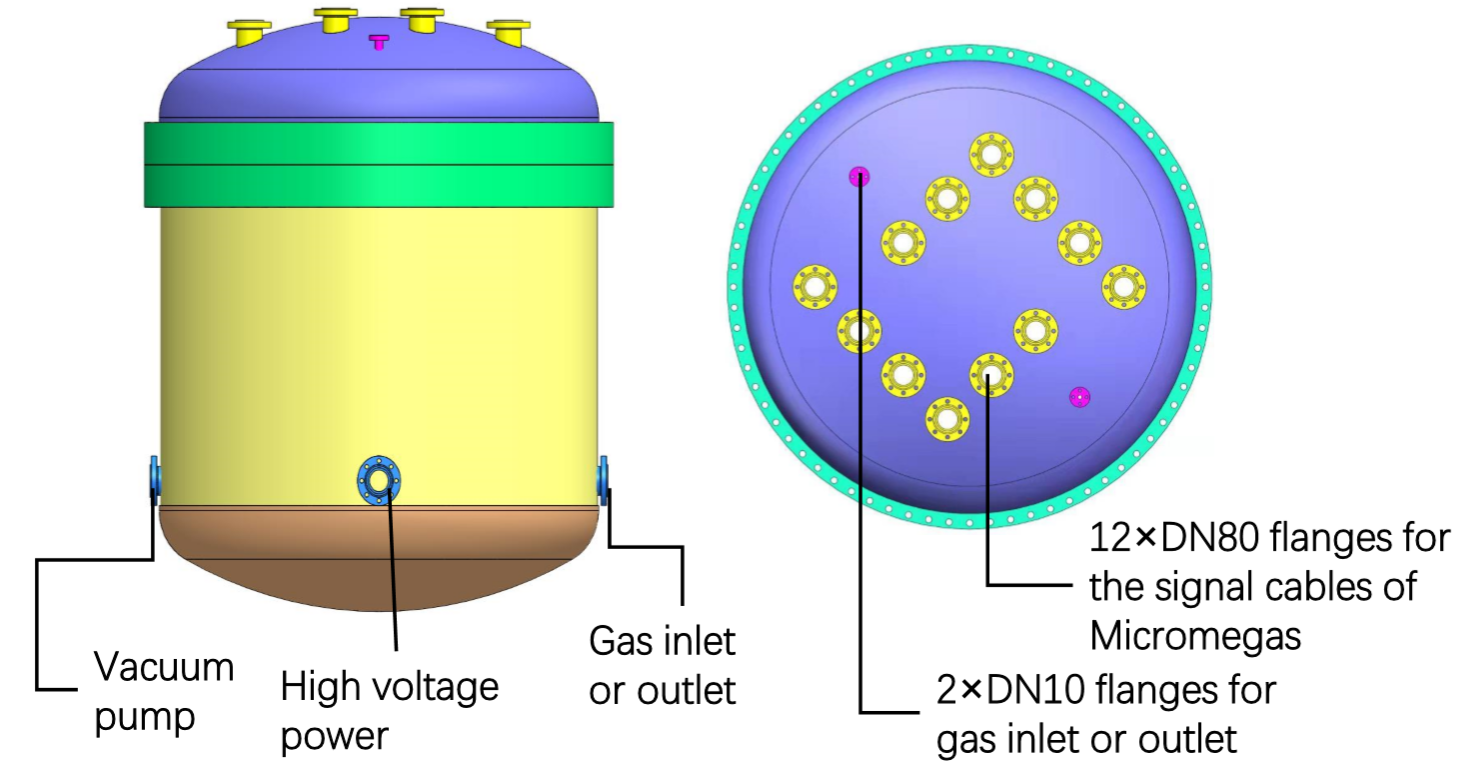}
    \caption{
    (Left): Drawing of  the pressure vessel of the detector with all the components.  There are four side ports on the lower bottom of the vessel  while only three are visible in the drawing. 
    (Right): Drawing of  the top flat flange of the detector with all the components. There are 12 DN-80 flanges for the outpout of the MMs signals, and 2 DN-10 flanges for gas inlet of circulation.
    }
    \label{fig:vessel}
  \end{center}
\end{figure}

As shown in Fig.~\ref{fig:vessel}, the SS pressure vessel includes two parts.
The dished head flange (Fig.~\ref{fig:vessel}(Right) is on the up side with a thickness of 1.6~cm to 1.8~cm and with a height of 59.7~cm, which could provide enough mechanical strength.
The cylindrical side wall is 1.2~cm thick.
The outer diameter of the cylindrical main body (Fig.~\ref{fig:vessel}(Left)) is 200~cm and the height is 127.9~cm (the height of the barrel itself is 120~cm, and the thickness of the cathode support on the bottom is 7.9~cm).
The inner diameter is 160~cm.
The top flange is held onto the vessel with 84 M27 bolts. The total height of the detector is 232.7~cm.

\begin{figure}[tb]
  \begin{center}
    \includegraphics[width=0.4\linewidth]{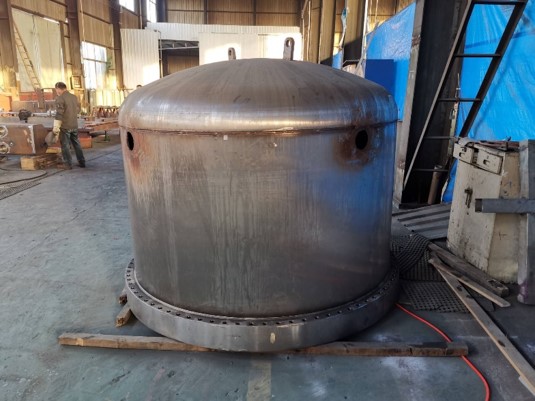}
    \includegraphics[width=0.4\linewidth]{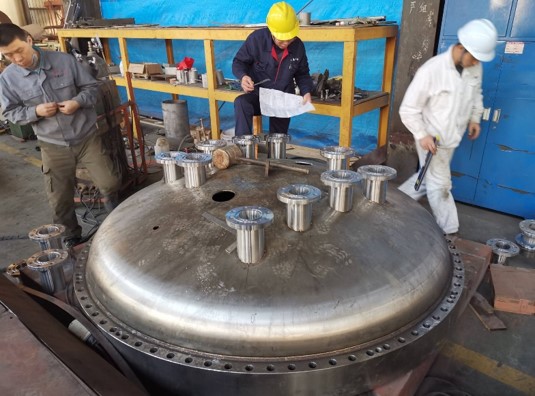}
    \caption{
    The pictures of the SS vessel of PandaX-III experiment,  the subsequent machining is going on at Shanghai.
    }
    \label{fig:vessel1}
  \end{center}
\end{figure}

On the top dished head flange, each of the 12 DN-80 flanges is used to hold from three to six Kapton extension cables for Micromegas signals readout and for bringing the high voltage bias.
The small DN-10 flange on the top is connected to a 1/4 inch SS pipe as gas outlet, besides a backup port.
Four DN-80 ports are at the lower part of the cylinder, besides a backup port, the other three are for the high voltage feedthrough of the cathode, gas inlet, and vacuum pumping.
All flanges are designed to withstand the required vacuum and high pressure conditions.
Expanded PTFE gaskets are used for all flange fittings, including the dished head flange. As shown in Fig.~\ref{fig:vessel1}, now the mechanical outsourcing processing of the vessel has been finished, and the subsequent machining is in progress in Shanghai.

\subsection{Charge readout plane}
\label{sec:readout}

\begin{figure}[tb]
  \begin{center}
    \includegraphics[width=0.6\linewidth]{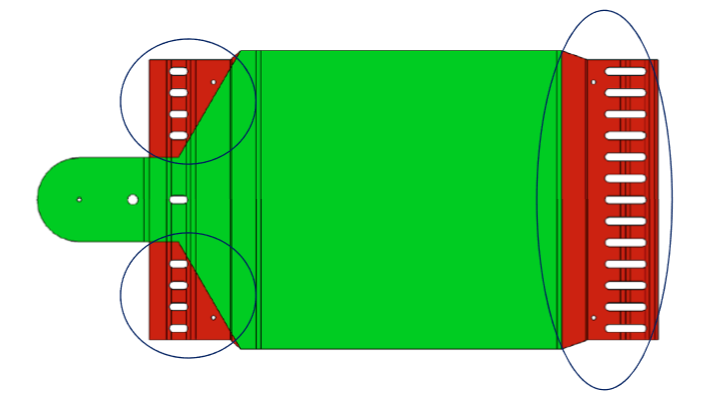}
    \includegraphics[width=0.3\linewidth]{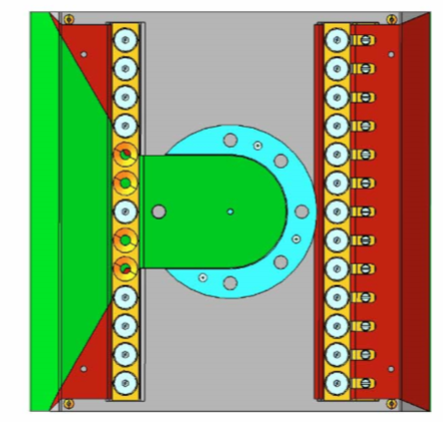}
    \caption{
    (Left): Drawing of the MMs film. the active volume is 20$\times$20~cm$^2$, the circuit of the signal is integrated on the tail of MMs, the two parts are marked in green. The red region without any circuit has a series of positioning holes for tensing. 
    (Right): Drawing of SR2M module, the MMs film is fixed on the back plate by tensing.
    }
    \label{fig:readout}
  \end{center}
\end{figure}

As shown in Fig.~\ref{fig:readout} (Left), the size of the Microbulk Micromegas is a square piece with 20.0~cm side length. The MMs is designed by Shanghai Jiao Tong University and Zarogoza University. The film of MMs is fabricated at CERN,
it is made of  Kapton and copper with a thickness of about 0.2~mm. There are 64 readout strips in each side (X and Y) with 3~mm piches. The area in green in Fig.~\ref{fig:readout} (Left) is the active region of the MMs,
and the flexible signal lines of the MMs (the tail) which are the green part of MMs  in Fig.~\ref{fig:readout} (Right).
The holes in the region in red without any circuit are used for tensing.
Finally the film is fixed by tensing on a copper back plate named SR2M~(Scalable Radio-pure Readout Modules) as shown in Fig.~\ref{fig:readout} (Right), 
so that we could arrange the charge readout plane with multi-modules mounted on a circular copper holding plate as shown in Fig.~\ref{fig:readout} (Left).

\begin{figure}[tb]
  \begin{center}
    \includegraphics[width=0.8\linewidth]{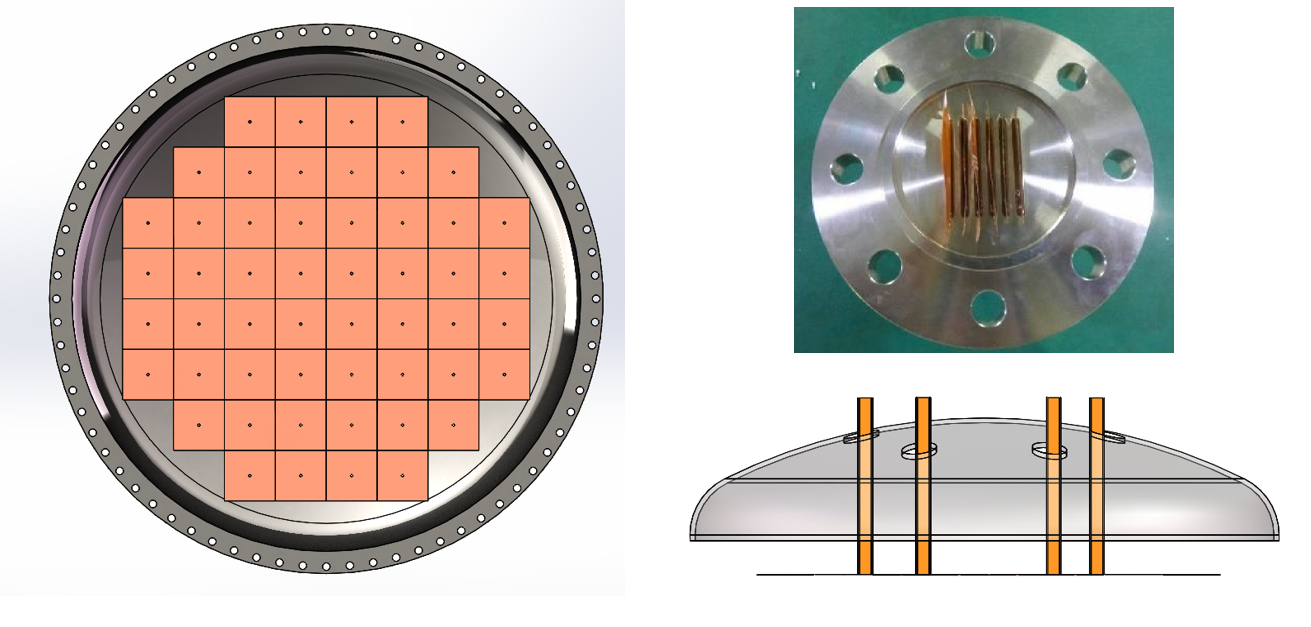}
    \caption{
    (Left): Drawing of the readout plane, which consists of 52 MMs with an active area 20$\times$20~cm$^2$. 
    (Top right): A picture of a DN-80 flange on which we glued 6 cables. Its tightness in 11 bar  has been tested.
    During assembly, the top flat flange and TPC hang from a blue frame as seen in the pictures (Bottom right): Schematic of the signal cables of the MMs, the cables will be glued on the flanges of DN-80 on the top flange.
    }
    \label{fig:readout1}
  \end{center}
\end{figure}

For the output of MMs signals, we designed the signal cables which are used to connect with the MMs inside the detector, the signal cables go through the DN80 fiange as shown in Fig.~\ref{fig:readout1} (Right bottom) to connect with the electronic system outside of the detector.
The signal cables will be glued in the flanges. By design, we need to glue 6 cables on the same flange for the output of all 52 MMs modules. 
We did a prototype with 6 cables glued as shown in Fig.~\ref{fig:readout1} (Top right) and tested its leakage in 11~bar xenon gas. The leakage rate is 100~g/yr for all the 52 signal cables. This result is acceptable for PandaX-III experiment. In addition, 
this type of signal cables glued in the flange has been used in a prototype detector~~\cite{Lin:2018mpd} for more than 3 years, no serious leakage is found, so the stability is reliable.

\subsection{Electric field cage}
\label{sec:FieldCage}

\begin{figure}[tb]
  \begin{center}
    \includegraphics[width=0.5\linewidth]{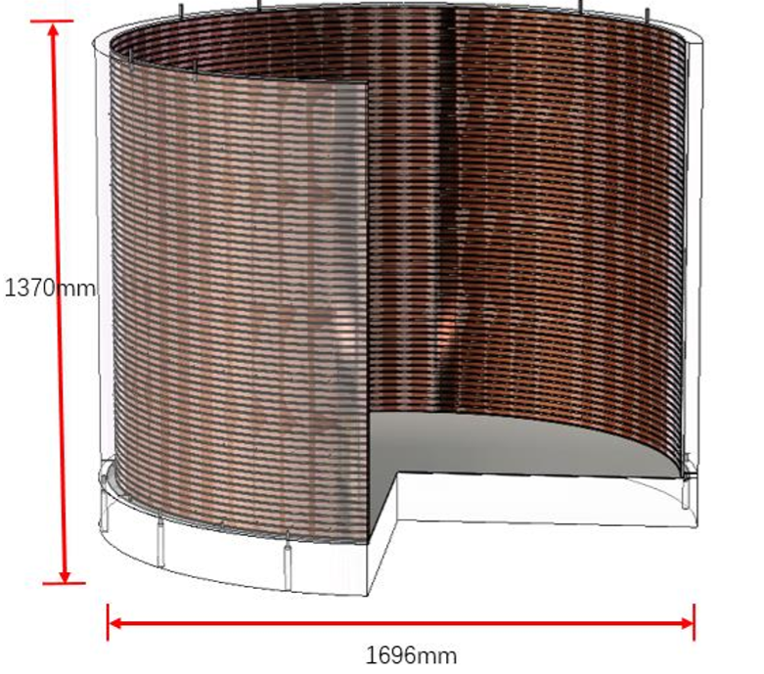}
     \includegraphics[width=0.35\linewidth]{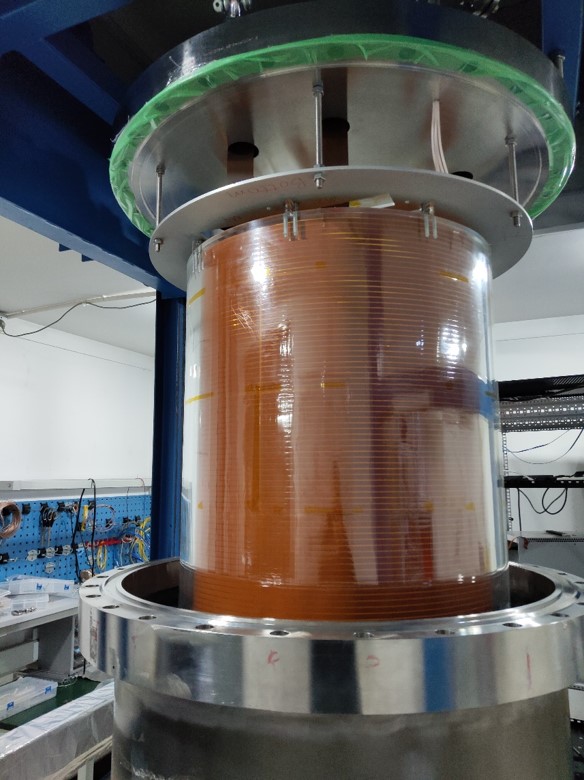}
    \caption{
   (Left): Schematic of the electric field shaping cage of the full detector of the PandaX-III experiment. It is made of segmented flexible PCBs, on which the coppers strips are connected by SMD resistors. The acrylic barrel is used to shape the electric field cage.
   (Right): A picture of the electric field cage of the prototype detector during assembly. Researchers are installing the electric field cage into the vessel of the prototype TPC to test the high voltage in nitrogen gas.
   }
    \label{fig:fieldCage}
  \end{center}
\end{figure}

As shown in the Fig.~\ref{fig:fieldCage} (Left), the electric field cage of the PandaX-III detector defines a cylindrical active volume of the TPC with a height of 120.0~cm, a diameter of 160.0~cm. 
It is made of kapton flexible PCB on which the copper strips serve as electrodes, the copper strips are connected with the SMD resistors of 1~G$\Omega$ to form the graded potentials and generate an uniform field orthogonal to them.
The kapton flexible PCB is supported by an acrylic barrel with a thickness of 4.8~cm, the cathode plane (SS plate with 0.5~cm thickness) on the bottom of the electric field cage is placed on an acrylic pedestal with a thickness of 17.0~cm  for protection and to avoid the discharge effect. 
In our TPC, electric field lines point from top to bottom and ionization electrons drift upward in the electric field cage. Compared with the previous design of the electric field cage~\cite{Chen:2016qcd,Chaiyabin:2017mis}, the novel one is much easier to assembly and has less material. 

\subsubsection{The electric field cage of the prototype TPC}
\label{sec:protpytpe field cage}
\begin{figure}[tb]
  \begin{center}
    \includegraphics[width=0.8\linewidth]{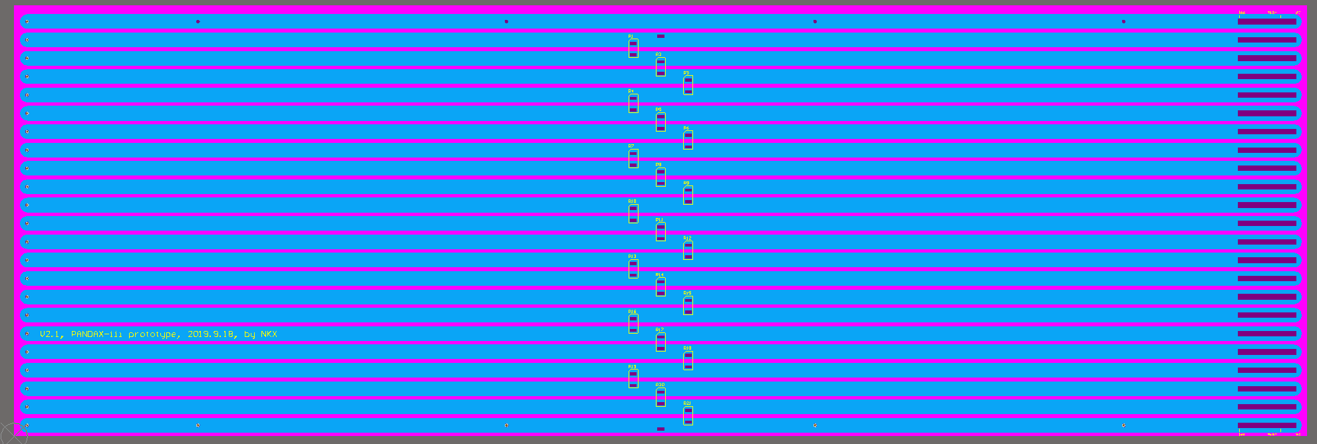}
    \caption{
   Schematic of the PCB of the electric field shaping cage. The blue bands are the copper strips, the pads of the resistors are following the three columns in the middle.  The pad of bar-shape on the right side of the copper strips is used to connect with the strips on the second PCB. 
   }
    \label{fig:pcb}
  \end{center}
\end{figure}

To study the performance of high pressure gaseous TPC and MMs, and also to optimize the design of all components of the detector, we have built a prototype TPC~\cite{Lin:2018mpd,Nim2019} at Shanghai Jiao Tong University. 
The TPC is single-ended, with the cathode at the bottom and charge readout plane on top. Firstly we designed an electric field cage for this prototype TPC as shown in Fig.~\ref{fig:fieldCage} (Right), its cylindrical shape has a diameter of 66~cm and a height of 78~cm (Fig.~\ref{fig:pcbtest} (Left)).
Due to the limitation of the PCB processing technology, it is not possible to fabricate a whole PCB board with a so large area, so we design a segmented PCB with a small area, then we splice small PCBs together for the full electric field cage. 
The current design is shown in Fig.~\ref{fig:pcb}, each piece of PCB is 70.4~cm by 23.5~cm (Fig.~\ref{fig:pcbtest}). There are 23 copper strips in each piece, so 21 resistors are needed to connect each other. The ends of each strip are smoothed to avoid the tip discharge effect.
The tags in the end of the strip are optimized for alignment and position adjustment, when we connect two segmented pieces. Finally, the electric field cage is spliced by 3$\times$4 pieces segmented PCBs as shown in Fig.~\ref{fig:fieldCage} (Right).

\subsubsection{Performance test}
\label{sec:Performance}
\begin{figure}[tb]
  \begin{center}
    \includegraphics[width=0.4\linewidth]{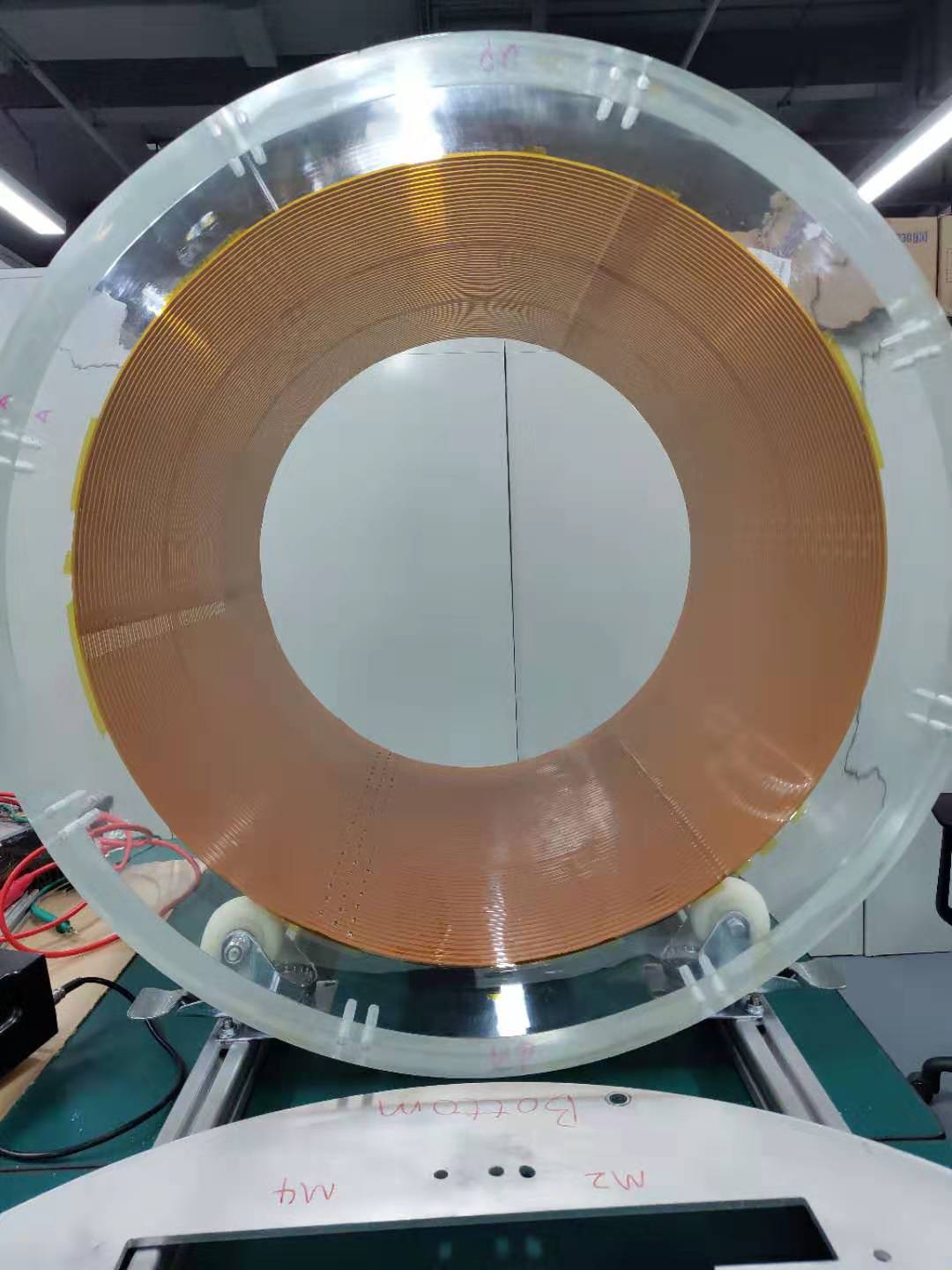}
    \includegraphics[width=0.4\linewidth]{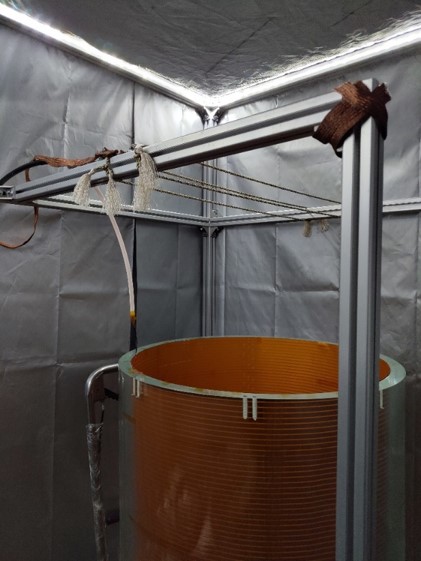}
    \caption{
   (Left): A picture of the barrel of the electric field cage and segmented PCBs glued on the barrel.
   (Right): A picture of the electric field cage of the prototype TPC during high voltage test. The cage is placed in a darkroom, the researchers are observing the location of  sparking .
   }
    \label{fig:pcbtest}
  \end{center}
\end{figure}

The test of the electric field cage of the prototype TPC was performed at Shanghai Jiao Tong University. Firstly  we tested the segmented PCB one by one in open air to get the high voltage threshold itself. The high voltage for the cathode is supplied by a Matsusada AU-100N3-L HV power supply.
Then we tested the whole electric field cage in a darkroom as shown in Fig.~\ref{fig:pcbtest} (Right) to observe the spark position. As the drifit length is 78~cm, a voltage of 78~kV is required to obtain an electric field of 1~kV/cm, which is our design goal~\cite{Chen:2016qcd}.
We did not find any sparks on the surface of the electric field cage with -80~kV applied on the cathode, but the boundary of the electric field cage attracted the curtain of the darkroom when we applied a higher voltage,  and then air in between was broken down.
Secondly, a similar test was performed with 1~bar nitrogen gas. We installed the electric field cage in the vessel of prototype detector (as shown in Fig.~\ref{fig:fieldCage} (Right)). We were flushing the vessel with nitrogen gas for 20~min before applying HV.
We found the nitrogen gas between the cathode and the vessel wall was broken down with -76~kV applied on the cathode.
As mentioned in the section~\ref{sec:FieldCage}, an acrylic pedestal is designed for the cathode of the final detector to avoid the discharge effect between the cathode and the vessel. For the test in the prototype TPC, it is expected to apply a higher voltage if an acrylic pedestal is used. 
And its uniformity will be further tested when we start the data taking.

\section{Conclusions and outlook}
\label{sec:outlook}
The PandaX-III collaboration is constructing a 140-kg scale gaseous Time Projection Chamber to search for Neutrinoless Double Beta Decay of $^{136}$Xe.
The gaseous TPC provides unique background suppression with tracking capability. 
The detector will be installed at China Jin-Ping underground Laboratory, a dry shielding system and a copper substrate are used to reduce the gamma background.
The half-life sensitivity to NLDBD will achieve 8.5$\times$10$^{25}$ years for an exposure of 3 years. 
In this paper we report the design of the dry shielding system firstly. And then we describe the components of the detector. The stainless steel pressure vessel allows us to operate the detector up to 10~bar.
We use 52 Microbulk Micromegas modules with strip channels of 3~mm for electron amplification and readout. Each Micromegas has an active area of 20$\times$20~cm$^2$. 
The signal cables glued on the top flanges are for the output of the Microbulk Micromegas signals. The electric field cage based on the flexible PCB is our first choice. A electric field cage of the prototype TPC has been built and tested.
Its performance achieved our goals, the design of the final electric field cage is going on. 

\bibliographystyle{unsrt}
\bibliography{draft_iprd2019}
\end{document}